\documentclass[11pt,letterpaper]{article}
\usepackage[letterpaper,margin=1in]{geometry}

\usepackage{amsmath,amssymb}
\usepackage{physics}
\usepackage{bm}
\usepackage{graphicx}
\usepackage{xcolor}
\usepackage{placeins}

\usepackage{authblk}

\usepackage[super,comma,sort&compress]{natbib}
\usepackage[font=footnotesize,labelfont=bf,labelsep=period]{caption}
\usepackage{titlesec}
\titleformat{\section}{\normalfont\large\bfseries}{\thesection}{1em}{}
\titleformat{\subsection}{\normalfont\normalsize\bfseries}{\thesubsection}{1em}{}

\usepackage{lineno}
\usepackage{setspace}

\usepackage[colorlinks,urlcolor=blue,citecolor=blue,linkcolor=blue]{hyperref}
\hypersetup{
  pdftitle={Programmable k-local Ising interactions and shallow optical Kolmogorov--Arnold networks through repeated data encounters},
  pdfauthor={Nikita Stroev and Natalia G. Berloff},
  pdfsubject={Repeated data encounters for programmable k-local Ising Hamiltonian evaluation and polynomial optical edge functions}
}

\title{\bfseries Programmable $k$-local Ising interactions and shallow optical
Kolmogorov--Arnold networks through repeated data encounters}

\author[1]{Nikita Stroev}
\author[2,*]{Natalia G. Berloff}
\affil[1]{Department of Physics of Complex Systems, Weizmann Institute of
Science, Rehovot 76100, Israel}
\affil[2]{Department of Applied Mathematics and Theoretical Physics,
University of Cambridge, UK}
\affil[*]{Corresponding author: \texttt{n.g.berloff@damtp.cam.ac.uk}}
\date{}

\newcommand{\kfourBenchRms}{0.1293}
\newcommand{\kfourBenchInf}{0.2747}
\newcommand{\kfourBenchMargin}{1.8125}
\newcommand{\kfourBenchWalshSector}{23}
\newcommand{\kfourBenchWalshMax}{7.63\times10^{-2}}
\newcommand{\kfourHalfRms}{0.0375}
\newcommand{\kfourHalfInf}{0.0847}
\newcommand{\kfourHalfMargin}{1.9514}
\newcommand{\kfourPatchRms}{1.0000}
\newcommand{\kfourWithinZeroIone}{144}
\newcommand{\kfourWithinZeroItwo}{151}
\newcommand{\kfourBootRmsMed}{0.1292}
\newcommand{\kfourBootMarginMed}{1.8098}

\begin{document}
\maketitle

\begin{abstract}
\noindent
Photonic processors are naturally suited to linear transformations, but independently programmable higher-order interactions usually require nonlinear media or a reduction to pairwise models. We introduce a repeated-encounter architecture that combines linear optical propagation with square-law detection to evaluate sparse and structured $k$-local Ising objectives. Each hyperedge is routed to a resolved channel, returned through the spin-dependent mask, and reconstructed from calibrated encounter-order signals. Within the stated architecture class, a $k$-spin Walsh term requires at least $\lceil k/2\rceil$ data encounters. A reciprocal rank-one recollection attains this bound for every finite order, allowing hyperedge identity, interaction order, and signed coupling to be programmed independently without quadratization ancillas or material optical nonlinearities. We test the $R=2$, $k=4$ member in a finite discrete-Fourier model of an ideal folded $4f$ relay. Reciprocal recollection produces the four-body response, whereas a fixed-patch control does not; configuration-level calibration measures the leakage caused by finite windows. Under signed-amplitude encoding, the same encounter hierarchy spans polynomial edge functions for shallow optical Kolmogorov--Arnold networks. The arbitrary-$k$ result is analytic; the finite Fourier calculation tests its two-encounter member and does not replace experimental validation.
\end{abstract}

\section{Introduction}\label{sec:introduction}

Ising machines couple a physical representation of an objective to dynamics or measurement feedback that searches for low-energy spin configurations \cite{mohseni2022ising,stroev2023analog}. Demonstrated platforms include coherent Ising machines based on optical parametric oscillators and laser networks \cite{yamamoto2017coherent,inagaki2016coherent,mcmahon2016fully}, polaritonic and photonic-lattice simulators \cite{berloff2017realizing,kalinin2020polaritonic,parto2020realizing}, injection-locked vertical-cavity surface-emitting-laser annealers \cite{zhang2025all}, monolithically integrated photonic Ising machines \cite{wu2025monolithically}, analog optical computers \cite{kalinin2023analog,kalinin2025analog}, and electronic simulated-bifurcation and probabilistic-bit engines \cite{goto2021high,camsari2017stochastic,camsari2019p}. Kerr-soliton and Hopfield-inspired architectures provide further optical realizations \cite{jin2025kerr,al2025programmable}.

Most widely programmable implementations still target pairwise couplings. Genuine higher-order terms arise in coding, constraint satisfaction, and many-body models \cite{lucas2014ising,monasson1997statistical,mezard2002analytic,mezard2003two,kempe2006complexity,pedretti2025solving}. Quadratization replaces such terms by pairwise couplings, ancillary variables, and penalty constraints \cite{boros2020compact,valiante2021computational}. Direct higher-order mappings can reduce this overhead \cite{bashar2023designing,bybee2023efficient}. Native three-body and higher-order CMOS, probabilistic-bit, and neuromorphic machines provide problem-level benchmarks \cite{su2023reconfigurable,nikhar2024reconfigurability,ahsan2026higher}; exact spin elimination converts large quadratic instances into smaller objectives with higher-order couplers \cite{berloff2025exact}; and gain-dissipative photonic networks have been proposed for direct discrete polynomial optimization \cite{stroev2019discrete}. Hardware that programs and evaluates higher-order terms directly is therefore useful independently of the search dynamics attached to it.

In photonic hardware, coefficient programmability and interaction order have largely appeared separately. Spatial photonic Ising machines use spatial light modulator (SLM) phase encoding and coherent Fourier-plane readout \cite{pierangeli2019large}; focal-plane division and related encoding schemes extend this architecture to arbitrary quadratic coupling matrices \cite{veraldi2025fully,sakellariou2025encoding}. Optical higher-order emulators use material nonlinearities, such as $\chi^{(2)}$ frequency conversion or Kerr parametric oscillation, and realize structured sectors with global rather than per-hyperedge coefficient control \cite{kumar2020large,kanao2021high}. Repeated data encoding in a parameter-dependent linear scattering system offers a third route: a nonlinear response to the encoded data without a nonlinear medium \cite{yildirim2024nonlinear,wanjura2024fully,wang2024large}. Here we use repeated encounters to construct independently weighted Ising interactions of arbitrary order and to determine the minimum optical depth. For continuous inputs, the same basis yields polynomial edge functions.

The central result is a tight encounter theorem. Each group of
interacting spins, or hyperedge, is routed to its own resolved optical
channel, where the light arriving after a single pass carries a signal
proportional to the sum of the spins in that group. We prove that no
scheme of this kind can do better per pass: when the incident light
carries no spin information, each modulator pixel depends on at most
one spin and at most linearly, propagation between encounters is fixed,
and detection is square-law, the detected signal can never couple more
spins jointly than twice the number of times the light has met the
data. A term coupling a given number of spins therefore requires at
least half that many data encounters, rounded up. An explicit
construction attains the bound. The selected light is returned as a
single uniform beam onto the corresponding spin-labelled macropixels, so each further
pass raises the attainable order by one; direct intensities recorded
after each pass supply the even powers of the spin sum,
phase-referenced interference between readouts of adjacent passes
supplies the odd powers, and calibrated weights assemble these channels
into the product of all spins in the group, with its programmed sign
and strength. A bank of a given encounter depth therefore programs
every interaction order up to twice that depth. Because each hyperedge
occupies its own selected return, the architecture is naturally suited
to sparse and structured objectives, and returns can be shared when the
spin groups and their routing labels are compatible.

We illustrate the theorem at its four-spin member, which requires two
data encounters, in a finite discrete-Fourier model of an ideal folded
4f relay. Each propagation step is an exact shifted two-dimensional
discrete Fourier transform, and the fold inverts both transverse
coordinates. The model retains finite selection windows, carrier
binning, ghost diffraction orders, and calibrated per-macropixel trims.
We compare reciprocal recollection with a fixed, data-independent return
using encounter scaling, the complete spin-product (Walsh) spectrum, a
patch-depth sweep, and reconstruction of the four-spin energy on all 16
configurations.
The complete finite-dimensional calculation is regenerated from source
and checked against a stored numerical record. A closed-form
algebraic solution, obtained with Cardano's formula, supplies an
optional optical seed for the grating depths that set the two
nonconstant weights. Replacing a binary spin by a bounded continuous
value gives the continuous-variable corollary: polynomial edge
functions for Kolmogorov--Arnold networks, of degree up to twice the
encounter depth, illustrated by recovering prescribed coefficients in a
fixed-projection ridge bank at depth two.

\begin{figure}[htbp]
\centering
\includegraphics[width=\textwidth,height=0.62\textheight,keepaspectratio]{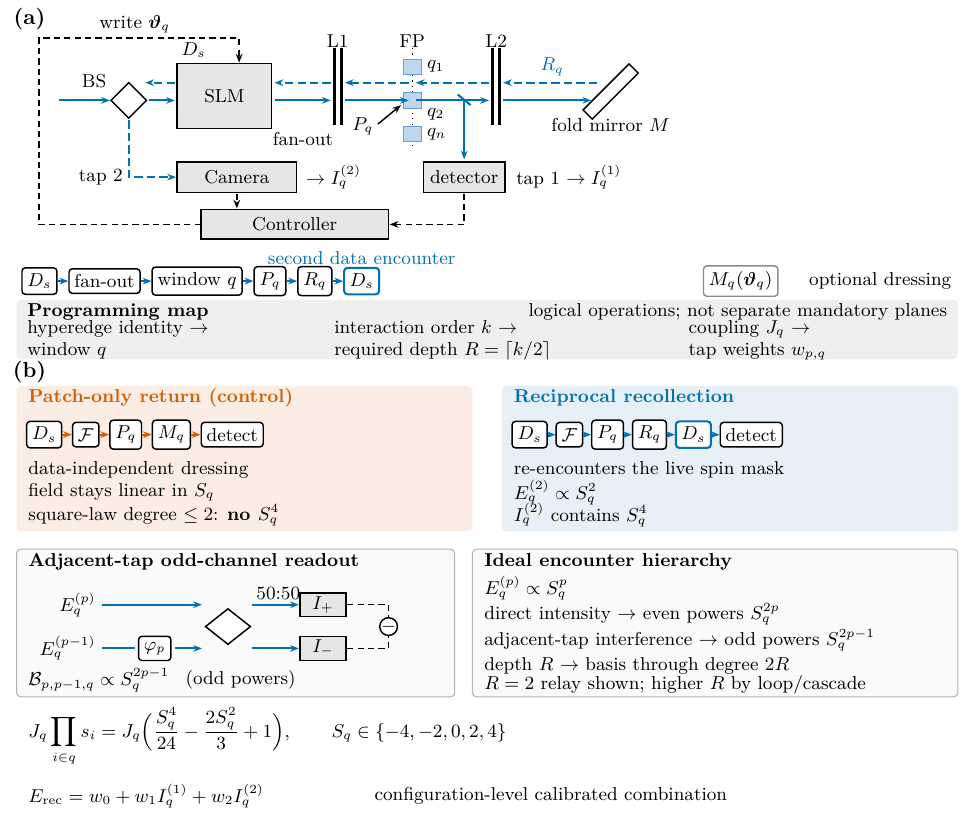}
\caption{\textbf{Encounter-resolved Hamiltonian programming in an ideal folded $4f$ relay.}
\textbf{(a)} The SLM applies the spin mask $D_s$ and a fan-out $\mathcal F$ that sends hyperedge $q$ to a resolved Fourier-plane window. The selector $P_q$ and reciprocal-recollection operator $R_q$ return the selected field to the corresponding spin-labelled replica macropixels; the first- and second-encounter signals are read as separate taps. In the numerical model, $R_q$ consists of the fold parity and the return-block blaze in Region B of Supplementary Table~S2. The controller writes raw phase controls $\boldsymbol{\vartheta}_q$, whose calibrated coordinates are $\boldsymbol{\theta}_q$. An optional dressing $M_q$ changes route gains without changing spin degree.
\textbf{(b)} A data-independent patch leaves the field at spin degree one and its square-law signal at degree at most two. Reciprocal recollection gives $\mathcal E_q^{(2)}\propto S_q^2$, so its intensity contains $S_q^4$. At depth $R$, rank-one recollection gives $\mathcal E_q^{(p)}\propto S_q^p$; direct intensities provide $S_q^{2p}$ and the balanced adjacent-tap observable $\mathcal B_{p,p-1,q}$ provides $S_q^{2p-1}$. For $R=2$, $E_{\mathrm{rec}}=w_0+w_1I_q^{(1)}+w_2I_q^{(2)}$ reconstructs the $k=4$ hyperedge polynomial.}
\label{fig:spim}
\end{figure}

\section{Results}\label{sec:results}
\subsection{Tight encounter theorem and constructive Hamiltonian programming for arbitrary $k$}
\label{sec:hyperedgeRep}

Let $s=(s_1,\ldots,s_N)\in\{\pm1\}^N$. An energy with interactions up to order $K$ is
\begin{equation}
H_{\le K}(s)=\sum_{k=1}^{K}\;\sum_{\substack{q\subseteq\{1,\ldots,N\}\\ |q|=k}}
J^{(k)}_q\prod_{i\in q}s_i.
\label{eq:kLocalDefinition}
\end{equation}
Fix an order $k$, let $\mathcal Q_k$ denote the represented $k$-spin hyperedges, write $J_q\equiv J_q^{(k)}$, and define
\begin{equation}
S_q=\sum_{i\in q}s_i\in\mathcal S_k,
\qquad
\mathcal S_k=\{-k,-k+2,\ldots,k\}.
\end{equation}
Since $(k-S_q)/2$ spins in $q$ are negative,
\begin{equation}
\prod_{i\in q}s_i=(-1)^{(k-S_q)/2}.
\label{eq:product-from-sum}
\end{equation}
These $k+1$ nodes define a unique polynomial of degree at most $k$; for the alternating values in Eq.~\eqref{eq:product-from-sum}, its degree is $k$ and its parity matches $k$ \cite{trefethen2019approximation,koekoek2010hypergeometric}. A convenient expression follows from the coefficient of $t^k$ in $\prod_{i\in q}(1+ts_i)$:
\begin{equation}
\prod_{i\in q}s_i=\Phi_k(S_q)=
[t^k](1+t)^{(k+S_q)/2}(1-t)^{(k-S_q)/2}
=\sum_{\substack{r=0\\r\equiv k\pmod 2}}^{k}c_r^{(k)}S_q^r.
\label{eq:phi}
\end{equation}
For example,
\begin{equation}
\Phi_4(S_q)=\frac{S_q^4}{24}-\frac{2S_q^2}{3}+1.
\label{eq:k4poly}
\end{equation}

The order-$k$ energy is reconstructed from parity-matched observables,
\begin{equation}
E_k^{\mathrm{rec}}(s)=\sum_{q\in\mathcal Q_k}
\left[w_{0,q}+\sum_{p=1}^{\lceil k/2\rceil}
w_{p,q}\mathcal O_{p,q}(s)\right].
\label{eq:measured-energy}
\end{equation}
For even $k$, $\mathcal O_{p,q}$ is a direct tap intensity; for odd
$k$, it is a phase-referenced difference between adjacent taps. The
weights are calibration constants, fixed once and independent of the
spin configuration. The relay generates the basis functions; their
affine combination, the balanced subtraction, and the controller
updates remain electronic by default, while the optically weighted
variant below encodes the two nonconstant $k=4$ weights in the return
optics. In either form the layer is a Hamiltonian evaluator, not an
end-to-end all-optical solver.

The synthesis problem is therefore reduced to producing the parity-matched powers in Eq.~\eqref{eq:phi}. The next theorem gives the minimum encounter depth and the construction that reaches it.

\subsubsection{Tight encounter theorem for arbitrary $k$}

\noindent\textbf{Theorem 1 (encounter-degree lower bound).}
Let the incident field be independent of the spin configuration $s$, and
call a data encounter one passage through a spin-dependent mask. Assume that
at every encounter the mask is diagonal in the optical modes and that each
diagonal entry is an affine function of at most one spin, with coefficients
independent of $s$; which spin a given mode addresses may differ between
modes and between encounters. Assume that propagation between encounters is
linear and data independent, that detection is square-law, and that the
reported energy is a data-independent affine combination of the detected
observables. If no detected path contains more than $R$ data encounters,
then every field component has multilinear spin degree at most $R$, every
observable has degree at most $2R$, and a $k$-spin Walsh term therefore
requires
\begin{equation}
R\ge\left\lceil\frac{k}{2}\right\rceil.
\label{eq:encounter-lower-bound}
\end{equation}

\noindent\textit{Proof.}
The incident field has multilinear degree zero, and each encounter
multiplies every contributing field term by one affine factor, raising its
degree by at most one; linear data-independent propagation preserves the
bound, and reduction to multilinear form with $s_i^2=1$ can only lower it.
A square-law signal is a product of two such fields and has degree at most
$2R$; a fixed local oscillator adds a cross term of degree at most $R$; a
data-independent affine combination raises neither bound. Since the
multilinear Walsh expansion is unique, no observable of degree below $k$
contains a degree-$k$ character.
\hfill$\square$

\noindent\textbf{Corollary 1 (rank-one construction attaining the bound).}
Write $D_s=\operatorname{diag}(s_i)$ for the live spin mask, $\mathcal C_q$
for collection of the data-plane field into window $q$, and
$\mathcal U_q^{(p)}$ for the return of tap $p$ onto the spin macropixels of
$q$. Under uniform illumination the first pass gives
$\mathcal E_q^{(1)}=\xi_{1,q}S_q$. Assume each return delivers its input as
a single uniform beam on the corresponding spin-labelled macropixels, so that every return, mask,
and collection cycle multiplies the channel field by a nonzero calibrated
constant times $S_q$. The field at tap $p$ is then a pure power,
\begin{equation}
\mathcal E_q^{(p)}=\xi_{p,q}S_q^{p},
\qquad \xi_{p,q}\neq0,
\qquad p=1,\ldots,R,
\label{eq:rankoneencounter}
\end{equation}
with $\xi_{p,q}$ the accumulated route gain.

The intensity of tap $p$ is $|\xi_{p,q}|^2S_q^{2p}$, so direct detection
supplies every even power through $S_q^{2R}$. Odd powers follow from
interfering two taps at a balanced detector,
\begin{equation}
I_{\pm,q}^{(p,p')}=
\left|\mathcal E_q^{(p)}
\pm e^{\mathrm i\varphi_{p,p',q}}\mathcal E_q^{(p')}\right|^2,
\qquad
\mathcal B_{p,p',q}=
\frac{I_{+,q}^{(p,p')}-I_{-,q}^{(p,p')}}{4}
\;\propto\;S_q^{\,p+p'},
\label{eq:tapinterference}
\end{equation}
with the beamsplitter normalization absorbed into calibration. Interfering
each tap with its predecessor, and a fixed data-independent reference field
with tap 1, yields $S_q^{2p-1}$ for $p=1,\ldots,R$; the analyzer phase
$\varphi_{p,p',q}$ must keep the selected cross term nonzero
(Supplementary Note~S1). Dividing each channel by its calibrated gain and
weighting the results by $J_q$ times the coefficients of $\Phi_k$ then
reproduces $J_q\prod_{i\in q}s_i$ exactly: even $k=2R$ uses the constant and
the even channels, odd $k=2R-1$ the odd channels alone. These weights exist
and are unique, because on the admissible sums the even basis
$1,S^2,\ldots,S^{2R}$ is a Vandermonde system in $S^2$ and the odd basis
reduces to the same system after factoring out one power of $S$. Rank-one
recollection, tap access, and phase-referenced odd readout therefore attain
the bound of Theorem~1, $R_{\min}=\lceil k/2\rceil$, and a bank of depth
$\lceil K/2\rceil$ supplies every order $k\le K$.
\hfill$\square$

In the ideal construction each channel depends only on $S_q$, so the
parity-reduced sum nodes fix its calibrated gain. Finite apertures can
distinguish configurations that share a sum; the basis is then fitted on
the full configuration domain,
\begin{equation}
M^{\rm cfg}_{\zeta p}=
\begin{cases}
1, & p=0,\\
\mathcal O_{p,q}(s^{(\zeta)}), & p=1,\ldots,\lceil k/2\rceil,
\end{cases}
\qquad
\widehat{\mathbf w}_q=
\arg\min_{\mathbf w_q}
\|\mathbf M^{\rm cfg}\mathbf w_q-\mathbf y\|_2^2,
\label{eq:configuration-matrix}
\end{equation}
with target $y_\zeta=J_q\prod_{i\in q}s_i^{(\zeta)}$. Exact synthesis is
possible precisely when $\mathbf y$ lies in the column space of the measured
channels, and stability is read from the singular spectrum after physically
meaningful column scaling. The generic fit has $2^k$ rows per inequivalent
hyperedge geometry; preserved permutation symmetry, or a reduced leakage
model, restores the parity-reduced problem. Supplementary Notes~S1 and S2
give the reference subtraction, weighting, phase coordinates, and the
configuration-level conditioning test.

\subsubsection{Reciprocal recollection, patch control, and overlapping returns}
\label{sec:overlap}
The theorem separates the two return routes compared numerically below. A
data-independent patch acts on the once-encoded field: it can rescale it and
add a coherent background, but it cannot raise the spin degree, so its
square-law signal is at most quadratic in $S_q$. Reciprocal recollection
instead returns the selected mode through the live spin mask, executing a
second data encounter; by Corollary~1 the data-dependent field becomes
$\xi_{2,q}S_q^2$, and the detected intensity acquires a quartic term,
\begin{equation}
I_q^{(\mathrm{patch})}
=\Gamma_{0,q}+\Gamma_{1,q}S_q+\Gamma_{2,q}S_q^2,
\qquad
I_q^{(\mathrm{RH})}
=\bar\Gamma_{0,q}+\bar\Gamma_{2,q}S_q^2+\bar\Gamma_{4,q}S_q^4,
\qquad
\bar\Gamma_{4,q}=|\xi_{2,q}|^2.
\label{eq:patch-vs-recollection}
\end{equation}
The linear patch term vanishes for an incoherent background or a
parity-symmetrized readout. Finite windows and macropixel errors add
lower-order Walsh components, but the four-spin sector remains excluded from
the patch route by Theorem~1.

Direct taps may be recorded by pick-offs, multiplexed ports, or repeated
exposures with the spin pattern held fixed; odd channels use coherent
interference of two taps or an equivalent phase-referenced reconstruction.
Higher encounter depth follows by repeating re-imaging, multiplication by
$D_s$, recollection, and tap acquisition.

The selector returns one hyperedge at a time. When several hyperedges are
served simultaneously, the return written for $q$ can leak into the
collection window of a different hyperedge $r$. Because every pixel carries
a single spin, this leakage is a sum of single-spin contributions, and it
vanishes for every configuration precisely when each per-spin overlap
between the return of $q$ and the collection of $r$ is zero. Uniform
recollection violates this for hyperedges that share spins: returning
$q=\{1,2\}$ into $r=\{2,3\}$ contributes $s_2S_q=1+s_1s_2$, a spurious
offset and pair coupling. Disjoint hyperedges can therefore share a return,
while overlapping hyperedges need scheduling or orthogonal labels. Let
$\mathcal G_\cap$ be the hyperedge-intersection graph and $d_i$ the number
of active hyperedges containing spin $i$; reusable label groups number at
least $\chi(\mathcal G_\cap)\ge\max_i d_i$, whereas duplicating spin
macropixels uses $\sum_q|q|$ incidence slots. The two counts represent
physical resource costs of different kinds.

\subsubsection{Two-encounter $k=4$ instantiation}

A four-spin term needs $R=2$. Write
\[
I_q^{(1)}=\Gamma_{0,q}^{(1)}+\Gamma_{2,q}^{(1)}S_q^2,
\qquad
I_q^{(2)}=\Gamma_{0,q}^{(2)}+\Gamma_{2,q}^{(2)}S_q^2
+\Gamma_{4,q}^{(2)}S_q^4.
\]
The reconstructed contribution is
\begin{equation}
E_{\mathrm{rec}}=w_0+w_1I_q^{(1)}+w_2I_q^{(2)}.
\label{eq:k4synth}
\end{equation}
Matching Eq.~\eqref{eq:k4poly} gives
$w_2=J_q/(24\Gamma_{4,q}^{(2)})$,
$w_1\Gamma_{2,q}^{(1)}+w_2\Gamma_{2,q}^{(2)}=-2J_q/3$, and
$w_0=J_q-w_1\Gamma_{0,q}^{(1)}-w_2\Gamma_{0,q}^{(2)}$.
The sign of $J_q$ is carried by the weights. The construction requires a nonzero quadratic first tap and quartic second tap; the measured column-space test in Eq.~\eqref{eq:configuration-matrix} is the finite-aperture criterion.

\paragraph{Optically weighted variant and Cardano seed.}
The two nonconstant weights can instead be encoded by a return-path grating. Balanced output ports carry their signs, while the grating depths set their magnitudes:
\[
\psi_q^{\mathrm{gr}}(\mathbf r_\perp)=
\alpha\cos(2\mathbf k_{\rm g}\!\cdot\!\mathbf r_\perp)
+\beta\sin(4\mathbf k_{\rm g}\!\cdot\!\mathbf r_\perp).
\]
The second diffraction order of the $\alpha$ grating and the first order of the $\beta$ grating meet in one output cell. With two cosine gratings their amplitudes are in quadrature and the mixed contribution cancels; the sine displacement rotates the second route into the detected quadrature. Using $\mathcal J_1(x)\simeq x/2$ and $\mathcal J_2(x)\simeq x^2/8$, let $a_{22}$ and $a_{42}$ be the quadratic gains of the $2\mathbf k_{\rm g}$ and $4\mathbf k_{\rm g}$ routes and $a_{44}$ the quartic collision gain. Then
\begin{equation}
G_1(\alpha,\beta)=a_{22}\frac{\alpha^2}{2}
+a_{42}\frac{\beta^2}{2},
\qquad
G_2^{\mathrm{bal}}(\alpha,\beta)=a_{44}\frac{\alpha^2\beta}{8}.
\label{eq:cardano-route-gains}
\end{equation}
For target coefficients $\tau_{2,q}=-2J_q/3$ and $\tau_{4,q}=J_q/24$,
\begin{equation}
G_1=|\tau_{2,q}|,
\qquad
G_2^{\mathrm{bal}}=|\tau_{4,q}|.
\label{eq:gainsys}
\end{equation}
Eliminating $\beta$ and setting $u=\alpha^2$ gives
\begin{equation}
\lambda_3u^3-2\lambda_2u^2+\lambda_0=0,
\qquad
\lambda_3=a_{22},\quad
\lambda_2=|\tau_{2,q}|,\quad
\lambda_0=\frac{64a_{42}\tau_{4,q}^2}{a_{44}^2}.
\label{eq:cardano-cubic}
\end{equation}
Within this leading-order, unconstrained model, a positive root exists if and only if
\begin{equation}
\lambda_0\le\frac{32\lambda_2^3}{27\lambda_3^2},
\qquad
54a_{42}a_{22}^2\tau_{4,q}^2
\le a_{44}^2|\tau_{2,q}|^3.
\label{eq:cardano-feasibility}
\end{equation}
For the four-spin ratio $|\tau_{2,q}|=16|\tau_{4,q}|$,
\begin{equation}
|J_q|\ge J_{\min}=
\frac{81a_{22}^2a_{42}}{256a_{44}^2},
\label{eq:cardano-jmin}
\end{equation}
while $J_q=0$ switches both routes off. Equation~\eqref{eq:cardano-jmin} is the feasibility threshold of this leading-order two-grating weighting map. The encounter construction and electronic weighting do not inherit it. The selected root must also lie inside the calibrated depth interval.

Define
\[
\mathsf P_{\rm C}=-\frac13\left(\frac{2\lambda_2}{\lambda_3}\right)^2,
\qquad
\mathsf Q_{\rm C}=\frac{\lambda_0}{\lambda_3}
-\frac{2}{27}\left(\frac{2\lambda_2}{\lambda_3}\right)^3.
\]
In the feasible three-real-root regime, the Cardano branches are
\begin{equation}
\begin{aligned}
u_b={}&\frac{2\lambda_2}{3\lambda_3}
+2\sqrt{-\frac{\mathsf P_{\rm C}}{3}}
\cos\!\left[
\frac13\arccos\!\left(
\frac{3\mathsf Q_{\rm C}}{2\mathsf P_{\rm C}}
\sqrt{-\frac{3}{\mathsf P_{\rm C}}}\right)
-\frac{2\pi b}{3}\right],\\
\alpha_b={}&\sqrt{u_b},
\qquad
\beta_b=\frac{8|\tau_{4,q}|}{a_{44}u_b},
\qquad b=0,1,2.
\end{aligned}
\label{eq:cardano-trig}
\end{equation}
We retain positive roots with $0.08\le\alpha_b,\beta_b\le0.40$~rad and choose the one minimizing $\max(\alpha_b,\beta_b)$.

\begin{figure}[htbp]
\centering
\includegraphics[width=\textwidth]{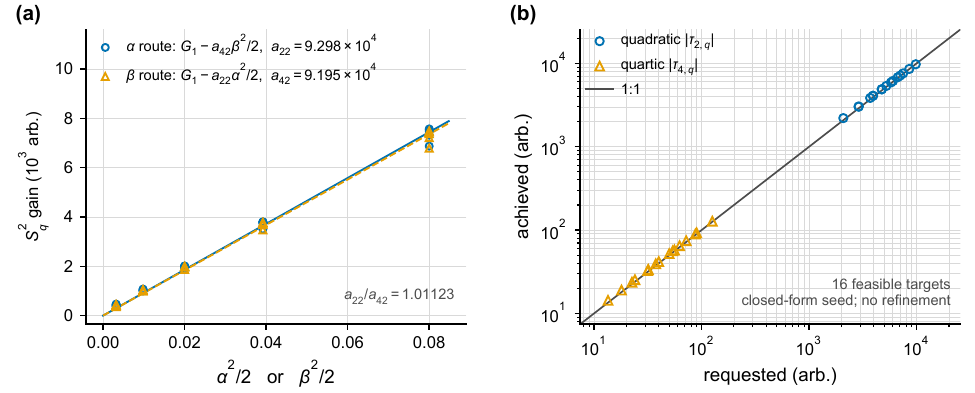}
\caption{\textbf{Route-gain calibration and Cardano seeding for the optically weighted $k=4$ variant.}
\textbf{(a)} Partial-residual fits isolate the quadratic gains of the $\alpha$ and $\beta$ routes.
\textbf{(b)} Gains obtained from Eq.~\eqref{eq:cardano-trig}, without refinement, versus the requested quadratic and quartic targets. Across 16 feasible targets, the errors are $0.41$--$5.95\%$ and $1.38$--$6.36\%$, respectively. The discrepancy combines finite-window leakage with neglected higher-order Bessel terms.
Both panels are numerical calibration results from the finite relay model.
Panel (b) uses the closed-form Cardano seed without iterative refinement.}
\label{fig:cardano}
\end{figure}

The simulated mixed gain scales as $\alpha^{1.97}\beta^{0.98}$ with $R_{\rm fit}^2=0.9999$, and the cosine/cosine quadrature control is suppressed by $8.6\times10^2$. The single-encounter gains satisfy $a_{22}/a_{42}=1.01$. Across the feasible targets in Fig.~\ref{fig:cardano}, the closed-form seed reproduces both requested gains within $6.4\%$ before refinement; the residual arises from finite-window leakage and the truncated Bessel expansion. Equation~\eqref{eq:k4synth} uses electronic weights and avoids this optical feasibility constraint.

\subsubsection{Finite-dimensional Fourier-relay verification}
\begin{figure}[htbp]
\centering
\includegraphics[width=0.98\textwidth]{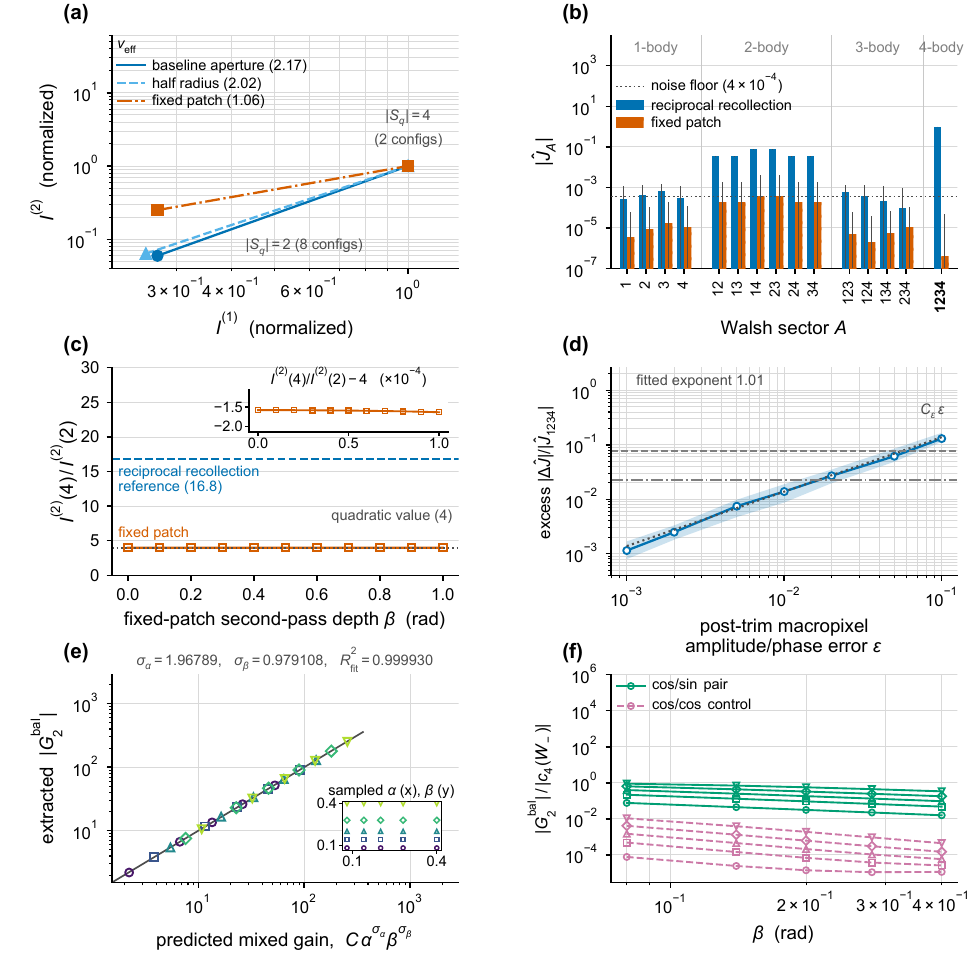}
\caption{\textbf{Fourier-relay verification for one $k=4$ hyperedge.}
\textbf{(a)} First- and second-tap window powers at $|S_q|=2$ and $4$, each normalized to its own $|S_q|=4$ mean. No background is subtracted: the routines return raw window sums. The legend gives the two-level exponent $\nu_{\rm eff}$.
\textbf{(b)} Complete nonconstant Walsh spectra over all 16 configurations, normalized to unit ideal four-body weight. Error bars are one bootstrap standard deviation for 20 synthetic repeats at $0.3\%$ multiplicative noise; the dotted line marks the $4\times10^{-4}$ four-body-sector bootstrap floor.
\textbf{(c)} The fixed-patch depth sweep remains at the quadratic ratio; reciprocal recollection is the reference.
\textbf{(d)} Excess Walsh content versus post-trim macropixel-error amplitude. The solid curve and band give the median and interquartile range over 40 seeds. A log--log regression of the median on the seven error levels gives fitted exponent $1.01$, annotated on the panel, and fitted linear-guide prefactor $C_\epsilon=1.379$. The dotted line is that guide, $C_\epsilon\epsilon$, drawn with the fitted prefactor; the symbol $C_\epsilon$ is a fitted prefactor and is unrelated to the interaction order $k$. The horizontal dash-dotted lines are the deterministic finite-window floors at the two first-pass radii.
\textbf{(e)} Numerically extracted balanced collision gain and the fitted law $G_2^{\rm bal}=C\alpha^{\sigma_\alpha}\beta^{\sigma_\beta}$, where $C$ is the fitted prefactor.
\textbf{(f)} The ratio $|G_2^{\rm bal}|/|c_4(W_-)|$ for the cosine/sine pair and the cosine/cosine quadrature control, where $c_4(W_-)$ is the four-spin Walsh coefficient extracted from the negative collision window $W_-$; each family contains one curve for each sampled $\alpha=0.08,0.14,0.20,0.28,0.40$~rad. Source Data contain all plotted values, fit ranges, and normalizations.
All panels are numerical results from the finite discrete-Fourier relay; no
experimental data are shown.}
\label{fig:walsh}
\end{figure}

We evaluate one $k=4$ hyperedge in the finite-dimensional Fourier relay
specified completely in Supplementary Table~S2: exact shifted
two-dimensional discrete Fourier transforms on a $1024\times1024$ grid,
with the written fan-out and return masks, finite windows, carrier binning,
ghost orders, fold parity, and calibrated per-macropixel trims. The control
replaces reciprocal recollection by the data-independent fixed patch. A
second setting halves only the first-pass selection radius, from $3.0$ to
$1.5$ FFT bins, leaving the second pass unchanged; we call it the
half-radius setting. Three diagnostics separate the routes
[Fig.~\ref{fig:walsh}(a--c)].

First, encounter scaling. Both routes give a first tap proportional to
$S_q^2$; an ideal second encounter makes the second tap quartic, while a
patch leaves it quadratic. The two-level exponent $\nu_{\rm eff}$ is the
log--log slope of the raw second-tap window power against the raw first-tap
window power, formed from the configuration-averaged levels $|S_q|=2$ and
$4$. The separate display normalization of each tap to its own $|S_q|=4$
mean does not change this slope. The exponent equals $2$ for an ideal second
encounter and $1$ for a patch. The simulations give $2.17$ at the baseline
first-pass radius, $2.02$ at the half-radius setting, and $1.06$ for the
patch.

Second, the complete Walsh spectrum. For any response $f(s)$ and sector
$\mathcal A\subseteq\{1,\ldots,4\}$, whose spins the subscript lists,
\begin{equation}
\widehat f_{\mathcal A}=2^{-4}
\sum_{s\in\{\pm1\}^4}f(s)\prod_{i\in\mathcal A}s_i,
\label{eq:walsh-definition}
\end{equation}
and $\widehat J_{\mathcal A}$ denotes these coefficients after normalizing
the response to unit ideal four-body weight. Reciprocal recollection gives
$\widehat J_{1234}=0.98$; the patch gives $4\times10^{-7}$, below the
$4\times10^{-4}$ bootstrap floor.

Third, the patch-depth sweep. Increasing the patch depth cannot raise the
spin degree: throughout the sweep the raw second-tap ratio between the
configuration-averaged $|S_q|=4$ and $|S_q|=2$ levels stays at the quadratic
value $4$ to relative accuracy $4.1\times10^{-5}$, while reciprocal
recollection gives $16.8$, near the ideal quartic value $16$.

Finite-window leakage is the leading deterministic error in this model: the
largest spurious normalized Walsh component is $7.8\times10^{-2}$ at the
baseline first-pass radius and $2.2\times10^{-2}$ at the half-radius
setting. The smaller selector contains $9/29=0.31$ as many detector samples
as the baseline selector. Because the intensity is not uniform across the
window, this sample-count ratio is not the collected-power fraction. Direct
integration of the first-pass field over the two selectors gives a retained
first-pass selected-window power ratio of $0.34$ for the all-plus
configuration and $0.32$ averaged over the 16 measured configurations,
spanning $0.04$ to $0.34$ across configurations. These are first-pass
window-selection ratios only, not a system throughput factor: they exclude the
second pass, detector efficiency, and every other loss in the chain. The
lower leakage should not be read as a net sensitivity improvement.
Post-trim macropixel amplitude and phase errors add excess Walsh content with
fitted slope $1.01$ versus error amplitude. A clean run matches all 40 stored arrays exactly in the stated
Python/NumPy environment, and the 24 numerical consistency checks reproduce
their reference outcomes. Supplementary Note~S3 gives the
finite-dimensional model, manifest, commands, and modeled error budget.

\subsubsection{Configuration-level reconstruction of the four-spin energy}
The finite-window channels are fitted on all 16 configurations with
Eq.~\eqref{eq:configuration-matrix}, taking $J_q=1$, uniform weights, and
no regularization; since fit and evaluation share the full domain, the
residual is pure representation error. Configurations with the same $S_q$
cannot be averaged. Among the six $S_q=0$ states, the relative ranges
$(\max I-\min I)/\overline I$ are $\kfourWithinZeroIone\%$ and
$\kfourWithinZeroItwo\%$ for the first and second taps, respectively,
because diagonal and adjacent sign patterns diffract differently. At the
baseline first-pass radius the reconstruction has RMS error
$\kfourBenchRms$, maximum error $\kfourBenchInf$, and a class margin, the
gap between the lowest reconstructed $+1$-target energy and the highest
$-1$-target energy, of $\kfourBenchMargin$ out of the ideal $2$. Halving
the first-pass radius improves all three: $\kfourHalfRms$,
$\kfourHalfInf$, and $\kfourHalfMargin$.

The residual concentrates in pair sectors, the largest being sector
$\kfourBenchWalshSector$ at $\kfourBenchWalshMax$. Two controls locate the
origin of the four-body information: the fixed-patch route gives RMS error
$\kfourPatchRms$ and no class separation, and the reciprocal first tap
alone gives the same null result. The four-body sector is therefore
carried by the second-encounter channel, not created by the affine fit.
The positive margin holds for this isolated hyperedge; residuals from many
terms can accumulate and reorder a global Hamiltonian.

\begin{figure}[htbp]
\centering
\includegraphics[width=\textwidth]{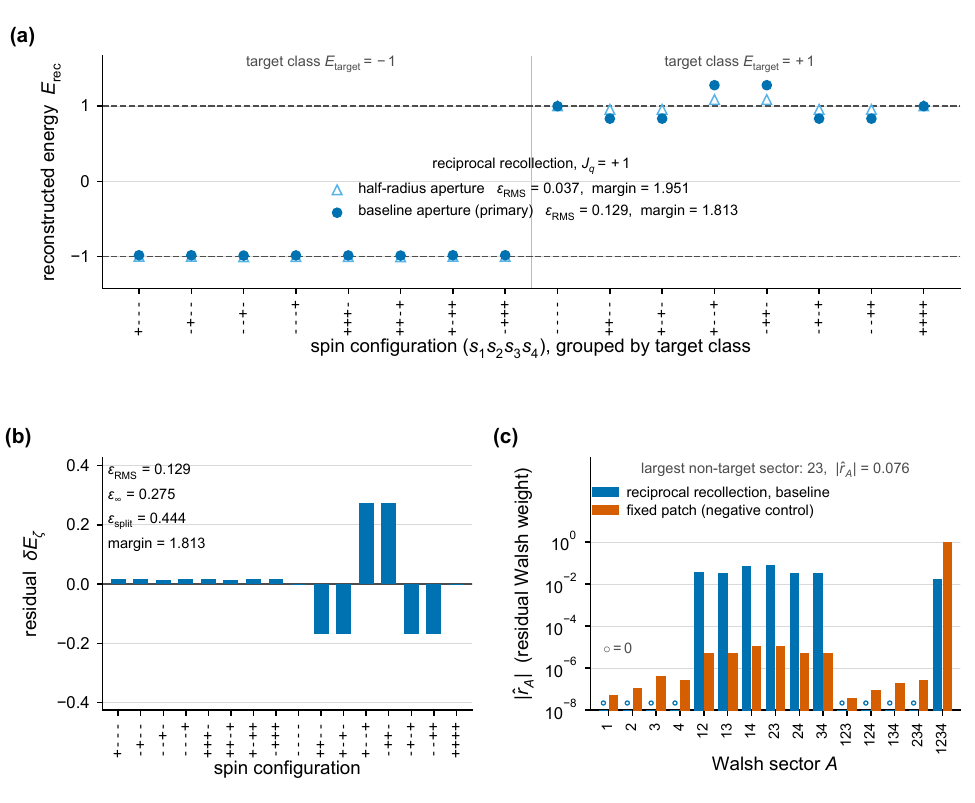}
\caption{\textbf{Configuration-level reconstruction of a four-spin energy from two encounter taps.}
\textbf{(a)} Reconstructed energies for all 16 configurations at the baseline and half-radius first-pass settings.
\textbf{(b)} Baseline residuals $\delta E_\zeta=E_{\rm rec}(s^{(\zeta)})-E_{\rm target}(s^{(\zeta)})$ and summary errors.
\textbf{(c)} Residual Walsh spectra for reciprocal recollection and the fixed-patch control. The unit four-body residual of the fixed patch is the unreconstructed target term; generated four-body content is absent on that route. Every configuration is fitted with uniform weight and no regularization.
Points show the 16 numerically evaluated spin configurations of the finite
relay model; the reconstructed values are obtained by the stated full-domain
affine fit.}
\label{fig:energy}
\end{figure}

A synthetic $0.3\%$ multiplicative-noise bootstrap, refitting all three
weights in every replicate, gives margin $\kfourBootMarginMed$ and RMS
error $\kfourBootRmsMed$; these quantify calibration sensitivity under
that noise model.

As a robustness check, Supplementary Note~S4 couples the surrogate
evaluator-error model to the single-spin Metropolis protocol specified there,
with its schedule, budget, success criterion, and seeds. At the operating
error level, the surrogate-evaluator arm recovers the planted target in
$194/200$ restarts, compared with $196/200$ for a noise-free
ideal-Hamiltonian control that removes both the deterministic pair-sector
contamination and the readout noise. The two arms share the instances,
schedule, and budget, and use independent restart streams. These are
descriptive counts over restarts of four fixed instances, not a solver,
hardware, or platform benchmark.

\subsection{Continuous-variable corollary: polynomial KAN edges}
\label{sec:opticalKAN}
A KAN layer applies a trainable univariate function $\phi_{ji}^{(\ell)}$ to
every edge $i\to j$ \cite{kan1961,liu2024kan},
\begin{equation}
x_j^{(\ell+1)}=\sum_{i}\phi_{ji}^{(\ell)}\!\left(x_i^{(\ell)}\right).
\label{eq:KANedge}
\end{equation}
Replacing the binary spin mask by the signed-amplitude transmittance
$D(z)=zD_1$, with $z$ a bounded coordinate normalized to $|z|\le1$, turns
the encounter construction into an edge-function generator: rank-one
recollection gives $\mathcal E_{ji}^{(p)}(z)=\xi_{p,ji}z^p$, direct
intensities supply $z^{2p}$, and the adjacent-tap observable of
Eq.~\eqref{eq:tapinterference}, with a fixed reference for $p=1$, supplies
$z^{2p-1}$. A depth-$R$ edge therefore realizes every real polynomial
\begin{equation}
\phi_{ji}(z)=\sum_{r=0}^{2R}g_{r,ji}z^r,
\qquad g_{r,ji}\in\mathbb R,
\label{eq:encounter-polynomial-edge}
\end{equation}
and the same encounter-degree counting, now applied to ordinary
polynomial degree, shows that a degree-$d$ edge requires
$\lceil d/2\rceil$ encounters; the construction attains this bound. An
unscaled coordinate $u$ with $|u|\le B$ enters as $z=u/B$, the factors
$B^r$ being absorbed into the calibrated coefficients.

The encoding fixes the function class. The strict degree-$2R$ identity
holds for transmittances affine in $z$, such as $D_0+zD_1$, whose
coefficients follow from a full-rank measured mixed-power basis; a
phase-only encoding such as $\exp[\mathrm i(\kappa z+b_{\rm ph})]$ instead
gives analytic channel responses, characterized on a working interval by
the rank, conditioning, and residual of the sampled response matrix.
Signed-amplitude operation uses a complex-amplitude modulator, dual rail,
or an equivalent interferometric front end.

\begin{figure}[htbp]
\centering
\includegraphics[width=\textwidth]{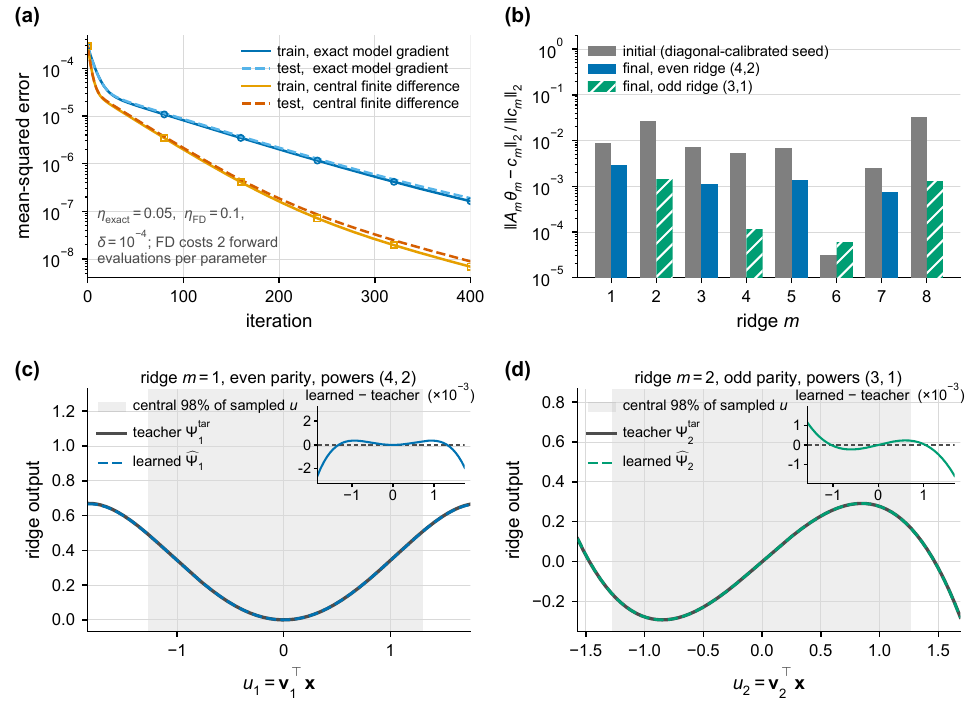}
\caption{\textbf{Coefficient recovery in the $R=2$ fixed-projection ridge bank.}
\textbf{(a)} Train and test MSE for exact model gradients and a central finite-difference update.
\textbf{(b)} Per-ridge relative coefficient error $\|\mathbf A_m\boldsymbol\theta_m-\mathbf c_m\|_2/\|\mathbf c_m\|_2$ before and after training. Both final series are the exact-gradient final parameters; the finite-difference errors are tabulated in the Source Data but are not plotted.
\textbf{(c,d)} Representative even and odd target functions and their learned reconstructions over the central $98\%$ of sampled raw projection values $u_m=\mathbf v_m^\top\mathbf x$. The target is $\Psi_m^{\rm tar}$ and the reconstruction is $\widehat{\Psi}_m$, drawn from the exact-gradient final parameters; physical signed-amplitude encoding uses $z_m=u_m/B_m$.
These are software device-model calculations for a fixed-projection
polynomial ridge bank; they do not use the finite Fourier-relay propagation
model or experimental optical data.}
\label{fig:kan_teacher}
\end{figure}

The numerical test is a fixed-projection ridge bank at $R=2$,
$y(\mathbf x)=\sum_{m=1}^{8}\Psi_m(\mathbf v_m^\top\mathbf x)$, with eight
fixed orthonormal projections $\mathbf v_m$ feeding alternating even
$(4,2)$ and odd $(3,1)$ polynomials $\Psi_m$. Each measured channel map
$\mathbf A_m$, which sends the two trained controls $\boldsymbol\theta_m$
to the ridge coefficients, stays fixed; only the 16 scalar controls are
trained toward the target coefficients $\mathbf c_m$. After 400 steps the
test MSE is $1.9\times10^{-7}$ with exact model gradients and
$8.9\times10^{-9}$ with central finite differences at a different learning
rate, two separate coefficient-recovery checks rather than a gradient
comparison. Figure~\ref{fig:kan_teacher} plots against the raw projections
$u_m=\mathbf v_m^\top\mathbf x$; the optical coordinate is $z_m=u_m/B_m$.
The checks reported here cover a single edge bank; stacking banks into a
multilayer network would require an explicitly specified interlayer detection
and re-encoding operation, which we do not characterize. Supplementary
Note~S5 gives the channel layout, coefficient targets, and software
checks (Supplementary Figs.~S2 and S3). Other photonic KAN designs realize
different edge-function families \cite{peng2025photonic}.

\FloatBarrier
\section{Discussion}

Repeated data encounters use linear propagation to generate a controlled
nonlinear dependence on encoded data
\cite{yildirim2024nonlinear,wanjura2024fully}. Under the assumptions of
Theorem~1, a $k$-spin Walsh term needs at least $\lceil k/2\rceil$
encounters, and rank-one recollection with tap access and phase-referenced
odd readout attains the bound. Spatial routing selects the hyperedge,
encounter depth sets its order, and calibration fixes its signed
coefficient. Focal-plane-division Ising machines already provide
programmable quadratic evaluation with digital feedback
\cite{veraldi2025fully}; nonlinear-optical and electronic machines realize
higher-order sectors through different physical mechanisms
\cite{kumar2020large,kanao2021high,su2023reconfigurable,nikhar2024reconfigurability,ahsan2026higher}.
The result here is an independently weighted arbitrary-order basis in
linear optics together with an encounter-depth bound.

The Fourier-relay calculation tests the $R=2$ construction in the finite
model of Supplementary Table~S2. Supplementary Note~S4 asks whether the
modeled error scale changes planted-target recovery under the single-spin
Metropolis protocol specified there, relative to a noise-free
ideal-Hamiltonian control. It is a descriptive check on four fixed instances
with independently seeded arms, not an inferential comparison of solver
performance. Neither calculation
provides a hardware time or energy to solution, a cross-platform ranking, or a
native-versus-quadratized performance comparison. Experimental transfer
also requires measured aberrations, apertures, SLM--camera response,
encounter loss, photon counts, detector noise, and phase drift.

Scaling is set by the number of resolved windows, selected-return
acquisitions, orthogonal labels, and calibration measurements. An objective
with $M$ represented hyperedges uses $M$ first-pass channels and up to $M$
selected returns. Edge-disjoint hyperedges can share a return; overlapping
hyperedges require the labels or duplicated pixels counted in
Sec.~\ref{sec:overlap}. A dense $k$-uniform objective needs
$\binom{N}{k}$ windows and per-spin fan-out $\binom{N-1}{k-1}$. The proposed
architecture is therefore intended for sparse and structured hypergraphs.
Higher encounter orders will require symmetry-preserving optical design,
transferable calibration, and compact models of finite-window leakage.

For continuous inputs, coordinate-selective routing gives canonical
polynomial KAN edges; the reported numerical calculation uses the
fixed-projection ridge-bank specialization. Experimental validation should
begin by calibrating reciprocal and patch returns on one four-spin geometry,
freezing the weights, and measuring all 16 configurations together with
photon counts, phase drift, and Walsh spectra. A later test should address
two overlapping hyperedges with controlled labels and phase-referenced
readout beyond $R=2$. Any hardware comparison with a quadratized baseline
would additionally require matched logical objectives, schedules, budgets,
and measured optical and controller costs.

\paragraph{Data availability.}
The numerical data and verification materials supporting this study are publicly available on Zenodo at \url{https://doi.org/10.5281/zenodo.21823681}.

\paragraph{Code availability.}
The code and configuration files used to generate and verify the results reported in this study are publicly available on Zenodo at \url{https://doi.org/10.5281/zenodo.21823681}.

\section*{Acknowledgements}
N.G.B. acknowledges support from the HORIZON EIC Pathfinder Challenges project HEISINGBERG (grant 101114978) and the EPSRC UK Multidisciplinary Centre for Neuromorphic Computing (grant UKRI982). Both authors thank the Weizmann--UK Make Connection grant (grant 142568) for support. N.G.B. used Anthropic's Claude for language editing, structural revision, and internal consistency checks. The authors independently verified all derivations, numerical results, and conclusions and take full responsibility for the manuscript.

\section*{Author contributions}
N.G.B. conceived the encounter-resolved $k$-local implementation, and N.S. conceived the KAN ridge-bank extension. Both authors developed the methodology and software, performed the analysis, interpreted the results, and wrote the manuscript. N.G.B. supervised the work.

\section*{Competing interests}
The authors declare no competing interests.

\bibliography{references_combined}
\end{document}